\documentclass[prl,superscriptaddress,twocolumn,amsmath,amssymb,showpacs]{revtex4}

\usepackage[T1]{fontenc}
\usepackage[latin1]{inputenc}
\usepackage{amsmath}
\usepackage{graphicx}
\usepackage{amssymb}

\usepackage{dcolumn}
\usepackage{bm}
\DeclareMathOperator{\erf}{erf}
\DeclareMathOperator{\sgn}{sgn}
\usepackage{upgreek}

\makeatletter

\begin{document}

\title{Entropic Splitter for Particle Separation}

\author{D. Reguera} \thanks{dreguera@ub.edu}
\affiliation{Departament de F\'isica Fonamental,
  Facultat de F\'isica, Universitat de Barcelona,
  Mart\'i i Franqu\`es 1, E-08028 Barcelona, Spain}

\author{A. Luque}
\affiliation{Departament de F\'isica Fonamental,
  Facultat de F\'isica, Universitat de Barcelona,
  Mart\'i i Franqu\`es 1, E-08028 Barcelona, Spain}

\author{P.S. Burada}
\affiliation{Max-Planck Institut f\"ur Physik komplexer Systeme,
N\"othnitzer Str. 38, D-01187 Dresden, Germany}

\author{G. Schmid}
\affiliation{Institut f\"ur Physik,
  Universit\"at Augsburg, Universit\"atsstr. 1,
  D-86135 Augsburg, Germany}

\author{J.M. Rub\'i}
\affiliation{Departament de F\'isica Fonamental,
  Facultat de F\'isica, Universitat de Barcelona,
  Mart\'i i Franqu\`es 1, E-08028 Barcelona, Spain}

\author{P. H\"anggi}
\affiliation{Institut f\"ur Physik,
  Universit\"at Augsburg, Universit\"atsstr. 1,
  D-86135 Augsburg, Germany}

\date{\today}


\begin{abstract}
We present a particle separation mechanism which induces motion of particles of different sizes in opposite directions. The mechanism is based on the combined action of a driving force and an entropic rectification of the Brownian fluctuations caused by the asymmetric form of the channel along which particles proceed. The entropic splitting effect shown could be controlled upon variation of the geometrical parameters of the channel and could be implemented in narrow channels and microfluidic devices.
\end{abstract}


\maketitle


Matter very often manifests as divided into very fine parts whose
nature dictate the overall properties of the system. Grains,
conglomerates or meso-structures dispersed in a liquid or liquid-like
phase are basic structures frequently found in many physico-chemical
and biological problems. The dissimilarity of these units, resulting
from random fractionation or self-assembling processes giving rise to
a very disparate distribution of sizes, originates a heterogeneous
response of the system which makes its typification difficult. To
obtain pure substances by separating wanted from unwanted elements
presents a main challenge in basic research and industrial processing, and in nanotechnology as well.

Particle separation techniques use the fact that the response of the particles
to external stimulus, such as gradients or fields, depends  on
their size. Filtering  particles of different size is
traditionally performed by means of centrifugal fractionation
\cite{separation}, phoretic forces \cite{DNA, DNA_nature,
  electrophoresis} or external fields \cite{sorting2}. By means of
these methods, the sorting of particles proceeds either by size
exclusion, as happens in a sieve, or by migration through the host
medium, a gel or porous media. In these cases all particles move in
the same direction but at different speeds. Novel separation
techniques based on flashing \cite{rousselet,Faucheux,Bader}, drift
\cite{drift_ratchet}, ``deterministic'' \cite{deterministic} and
geometric \cite{sorting,eichhorn} ratchets have also been proposed for sorting \cite{RMP-hanggi}.

Here we present a novel splitting mechanism that separates particles
in different directions by purely entropic means. The working
principle relies on the combined action of a static force and an
entropic rectification \cite{Kosinska, Schmid, chemphys, Zitserman}. Small particles follow the force whereas the
motion of big particles is rectified to proceed in the opposite
direction resulting in a faster splitting. This mechanism could be
readily implemented in microchannels or microfluidic systems. The
geometry of the channel can be tuned to be very selective leading to
an efficient separation of particles of very similar radius.

\begin{figure}[t]
\centering
\includegraphics[width=0.45\textwidth]{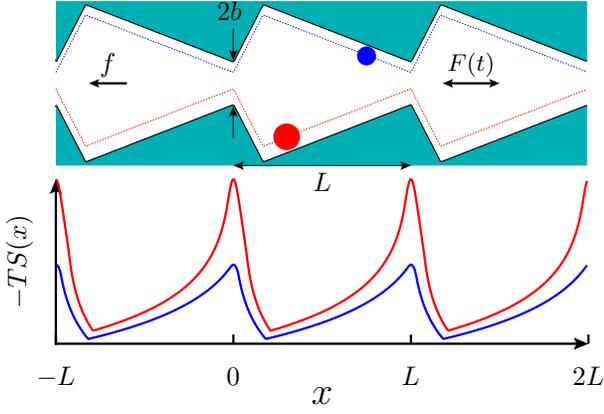}
\caption{(top) Schematic illustration of the two-dimensional
  channel confining the motion of the Brownian particles.
  The structure is defined by Eq.~\eqref{eq:channel}.
  Brownian particles are driven by a constant force $\vec{f}$ and a square wave
  force $\vec{F}(t)$ along the longitudinal direction. The values of the chosen parameters defined in Eq.~\eqref{eq:channel} are: $m_{1}=2$, $m_{2}=0.4$, $b/L=0.1$. The dashed lines represent the limit for the positions of the center of the particles within the channel. (bottom) Effective entropic potential for the two particle sizes depicted above.}
  \label{fig:channel}
\end{figure}

The Brownian motion of particles in confined geometries  exhibits a very rich and striking phenomenology
\cite{Reguera_PRE,Reguera_PRL,recti,Burada_PRE,Burada_PRL,Burada_EPL}. The
effects of confinement can be described by means of an effective
entropic potential resulting from the variation of the space
accessible to particles along the transport  direction
\cite{Jacobs,Zwanzig,Reguera_PRE}. The height of the entropic barrier
associated to the bottlenecks depends on particle radius. In an
asymmetric channel, the resulting entropic potential becomes
asymmetric, and rectification of a zero-mean oscillating force can
occur. The strength of this rectification depends on the particle radius
and is stronger for large particles (for which the entropic barrier is larger, see Fig. \ref{fig:channel}). Thus, in the presence of a small static force directed in the opposite direction of rectification, it is possible to separate particles of different sizes. The trajectories of small particles mainly follow the force, whereas large particles  move in the opposite direction. To illustrate this effect we have chosen the geometry depicted in Fig. \ref{fig:channel}.

The dynamics of a Brownian particle in a 2D channel as the one depicted in Fig.~\ref{fig:channel}, driven by a static force
$f$ and an
oscillating (square wave) force $F(t)$, both applied along the principal axis of the channel,
can be described by means of the Langevin equation which, in the
overdamped limit, reads
\begin{equation}
  \label{eq:langevin}
  \gamma \, \frac{\mathrm{d}\vec{r}}{\mathrm{d} t} = - \left( f+ F(t)\right)\vec{e_x} +\sqrt{\gamma \, k_{\mathrm{B}}T}\, \vec{\xi}(t)\, ,
\end{equation}
where $\vec{r}$ denotes the position of the particle, $\gamma$ is the
friction coefficient, $\vec{e_x}$ is
the unit vector along $x$-direction, and
$\vec{\xi}(t)$ is a Gaussian white noise with zero mean which
obeys the fluctuation-dissipation relation
$\langle \xi_{i}(t)\,\xi_{j}(t') \rangle = 2\, \delta_{ij}\,
\delta(t - t')$ for $i,j = x,y$. The explicit form of the oscillating driving force is $F(t) = A\, \sgn \left[\sin( \Omega t )\right]$ where $A$ is the amplitude, $\sgn[t]$ represents the sign function, and $\Omega$ is the driving frequency. This choice does not represent a restriction on the particle splitting effect which could also be observed for a sinusoidal driving force.

The Langevin equation \eqref{eq:langevin} must be solved by imposing vanishing outflow
at the walls of the structure. For the 2D structure depicted in
Fig.~\ref{fig:channel}, the walls are defined by
\begin{eqnarray}
y_{\mathrm{u}}(x) = \left\{ \begin{array}{ll}
   b + m_{1} \bar{x} &\mbox{ if $\bar{x} < c$} \\
   b + m_{2}(L -\bar{x})  &\mbox{ otherwise}
       \end{array} \right.
 \label{eq:channel}
\end{eqnarray}
where $ y_{\mathrm{u}}(x)$ and $ y_{\mathrm{l}}(x)= -y_{\mathrm{u}}(x)$ correspond to the upper and lower
 boundary functions,
respectively, $b$ is the half-width of the bottleneck, $m_{1}$ and $m_{2}$
are the slopes of the walls, $L$ is the periodicity of the channel, $c=L m_{2}/(m_{1}+m_{2})$ indicates
the location of the point of maximum width, and $\bar{x}= x \; \mathrm{mod}\;L$ is the modulo function (to create a
periodic structure) cf. Fig.~\ref{fig:channel}.

For a hard particle of radius $r$  inside the channel, the space available for its center is restricted
by a distance $r$ from the walls, reading

\begin{eqnarray}
        \label{eq:widthfunctions}
    w_{\mathrm{u}}(x) = \left\{ \begin{array}{ll}
		-\sqrt{r^2 - \bar{x}^2} + b, &\mbox{$0 \leq \bar{x} < o_{p}$} \\
		b + m_{1} \bar{x}- r\sqrt{(1 + m_{1}^2)},
			    &\mbox{$o_{p} \leq \bar{x} < c_{p}$}\\
		b + m_{2}(L -\bar{x}) - r\sqrt{(1 + m_{2}^2)},
	  &\mbox{$c_{p} \leq \bar{x} < L_{p}$}\\
		      -\sqrt{r^2 - (\bar{x} - L)^2} + b,  &\mbox{$L_{p} \leq \bar{x} < L$}\\
\end{array} \right.
\end{eqnarray}
where $o_{p}= r m_{1}/\sqrt{(1 + m_{1}^2)}$, $L_{p} = L -r m_{2}/\sqrt{(1 + m_{2}^2)}$, and
$c_{p}= c+ r\left(\sqrt{(1 + m_{1}^2)}-\sqrt{(1 + m_{2}^2)}\right)/(m_{1}+m_{2})$. The lower parallel curve is just $w_{\mathrm{l}}(x)=-w_{\mathrm {u}}(x)$.
Consequently, $2\,w(x) = w_{\mathrm {u}}(x) - w_{\mathrm{l}}(x)$
gives the local width of the structure accessible for the center of a hard particle of radius $r$.
Assuming throughout a dilute particle density and a strong viscous, low Reynold number dynamics, all
relevant hydrodynamic wall-particle interactions are small and of repulsive character \cite{Fuchs}.
Effectively this results in a slightly increased effective particle radius; i.e.,  $r_{eff} \gtrsim r$.
Such hydrodynamic interactions thus result in somewhat larger average transport velocities, note Figs. 2a, b below.

Also the friction depends on size, and using Stokes' law as an approximation one obtains $\gamma=\gamma_{0} r/b$ and $D=D_{0}b/r$,
where  $\gamma_{0}$ and $D_{0} = k_{\mathrm{B}}T/\gamma_{0}$ are the friction and diffusion coefficients of a particle of radius equal
to the bottleneck half-width $b$. To mimic the particular case of DNA electrophoreses, we will consider that the
forces depend linearly on the radius of the particles. Specifically,
we set
\begin{align}
  \label{eq:forcescaling}
  f=f_0 r/b \text{ and } A=F_0 r/b\, ,
\end{align}
where $f_0$ and $F_0$ are the strengths
of the static and periodic forces for a particle of radius $r$.

For the sake of a dimensionless description, we scale all variables  using
three characteristic parameters: the
characteristic length $L$, energy $k_{\mathrm{B}}T$, and diffusion coefficient $D_{0}$. Particularly, ${\tilde{x}} = x/L$, $\tilde{y} =
y/L$, ${\tilde {b}} = b/L$,
${\tilde {w}}_{\mathrm{l}}= w_{\mathrm{l}}/L = - {\tilde {w}}_{\mathrm{u}}$;  ${\tilde {t}} = t/\tau$ and $\tilde{\Omega} = \Omega \tau$, where
$\tau={L}^2/D_{0}$ is the
characteristic diffusion time. The scaled forces are:
${\tilde {f}} = fL/k_{\mathrm{B}}T$
and ${\tilde {F}(\tilde {t})} = F(t)L/k_{\mathrm{B}}T$.
In the following we shall omit the tilde symbols.
In dimensionless form the Langevin equation~\eqref{eq:langevin} reads:
\begin{align}
  \label{eq:dllangevin}
  \frac{\mathrm{d}\vec{r}}{\mathrm{d} t} & =
  -\left(f_0+ F_0 \sgn[\sin(\Omega t)]\right) \vec{e_x} +\sqrt{b/r}\, \vec{\xi}(t)\,.
\end{align}

The description of this system alternatively can be given by
the concept of an entropic potential \cite{Reguera_PRE} and the
corresponding Fick-Jacobs (FJ) equation \cite{Reguera_PRE,Reguera_PRL,Jacobs,Zwanzig}
\begin{equation}
\label{eq:fj}
\frac{\partial P(x,t)}{\partial t} =
     \frac{\partial}{\partial x}\bigg\{D(x)\left(\frac{\partial P}{\partial x} \, +
     V^{\prime}(x)\, P \right)\bigg\} \, ,
\end{equation}
where
\begin{equation}
\label{eq:effpotential}
V(x) = U-TS(x)= \left[F(t)+f\right]x - \ln\left[2\,w(x)\right],
\end{equation}
is the free energy including an entropic contribution $TS(x)=\ln[2\,w(x)]$, $D(x)=b/\{r[1+w'(x)^{2}]^{1/3}\}$ is the position-dependent diffusion coefficient, and the prime refers to the derivative with respect to $x$. This approximation is expected to be very accurate for bias strengths $|f_0| < 1$,  amplitudes $F_0 < 1$ and for $w'(x)^{2}\ll 1$ \cite{Burada_PRE}.

\begin{figure}[t]
\includegraphics[width=0.4\textwidth]{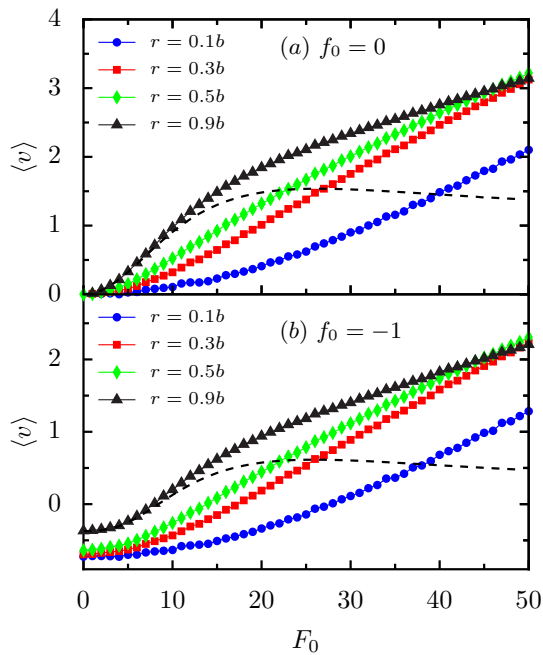}
 \caption{
 Average current {\itshape vs.} the amplitude of the periodic forcing $F_0$ in the adiabatic limit ( $\Omega=\Omega_0=\pi/10$) for particles of
different radius, for a channel with $m_{1}=2$, $m_{2}=0.4$. The solid lines indicate the results of the simulations, whereas the dashed line plots the prediction according to the FJ equation for $r=0.9b$. For $f_0<0$ small particles go
to the left, whereas big particles move to the right. The critical size that divides these two behaviors can be tuned
by the value of $F_{0}$.}
  \label{fig:current_simu}
\end{figure}

In the adiabatic limit, the average velocity can be calculated as
\begin{equation}
\langle v\rangle= \frac{J(F_0)+J(-F_0)}{2}
\label{velocity}
\end{equation}
where the current $J(F_0)$ is given by Stratonovich's formula \cite{stratonovich,HTB,Reimann}.
\begin{equation}
J(F_0)=\frac{1- e^{-(F_0+f_0)r/b}}{\int_{x_0}^{x_0+1} \, \mathrm{d}z \frac{1}{D(z)}\, e^{V(z)} \int_{z-1}^{z} \, \mathrm{d}x \,e^{-V(x)} }.
\label{stratonovich}
\end{equation}
Note that for very large driving strengths $F_0 \gg 1$, while keeping a finite bias strength $|f_0|$, i.e.
far beyond the regime of validity of the Fick-Jacobs approximation, the average velocity approaches the deterministic behavior \cite{Schmid,Reguera_PRL}. In this limit we then find that $J(F_{0})\approx - J(-F_{0})$ so that an asymptotic vanishing net velocity $<v>$ emerges. Note that this asymptotic regime of very large  amplitude strength $F_0$ is not yet reached in Fig.  \ref{fig:current_simu}.


\begin{figure}[t]
  \centering
  \includegraphics{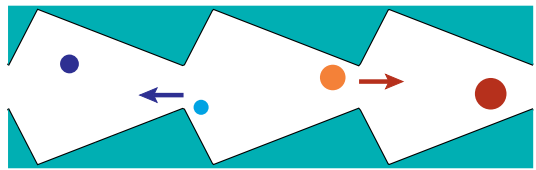}
  \caption{Schematic illustration of the functioning of the entropic splitter. A mixture of particles of two different radii is placed initially in the center. Under the combined presence of the static and periodic forcing, large particles move towards the right, whereas small particles follow the bias, i.e. are moving towards the left.}
  \label{fig:splitter}
\end{figure}

Fig. \ref{fig:current_simu} plots the average current {\itshape vs.} the amplitude of the periodic forcing $F_0$, in the adiabatic limit
($\Omega=\pi/10$), for particles of different radius.
In absence of static forcing, this asymmetric channel rectifies the oscillating force
giving rise to a net positive velocity, i.e. all particles move
towards the right (see Fig. \ref{fig:current_simu}$a$).
Its magnitude depends on the strength of the rectification which is
more intense the larger is the particle radius.
If we apply a small static force in the negative direction (see Fig. \ref{fig:current_simu}$b$),
for intermediate periodic forcing strengths $F_{0}$, particles larger than a given threshold radius move to the right,
whereas particles smaller than that move to the left.
In this way, one can separate particles of different radii and make them move in opposite directions.
The splitting effect is illustrated schematically with Fig. \ref{fig:splitter}.


Fig. \ref{fig:current_simu} also represents the results for the average velocity obtained from the FJ equation for the case $r=0.9b$. The agreement with the simulation results is very good at small values of the force (for $F_{0} < 10$). At larger values of $F_0$, the agreement is better when the radius increases or equivalently when entropic effects are more pronounced. In the domain of validity of the FJ description the corresponding applied forces already yield a significant entropic splitting.

\begin{figure}[t]
  \centering
  \includegraphics[width=0.45\textwidth,height=0.3\textwidth]{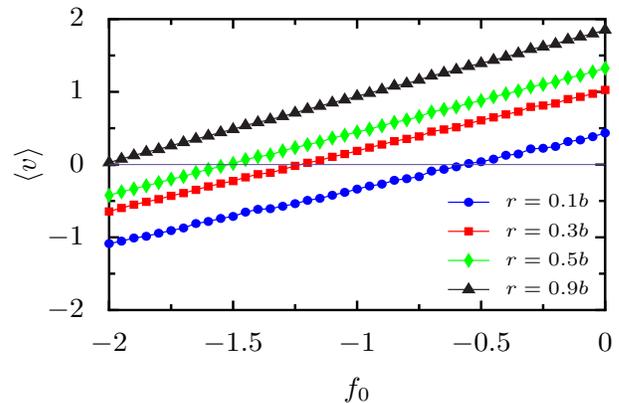}
  \caption{
  Average current {\itshape vs.} the value of the static force $f_0$ in the adiabatic limit ($\Omega=\pi/10$) for particles of
different radius, for the channel plotted in Fig. \ref{fig:channel} with $F_0=20$. In all cases the velocity increases linearly with $f_0$. As $f_0$ gets more negative, the particles invert their velocity and start to move to the left.  The critical force that determines this velocity inversion depends on the radius $r$ and gets progressively larger (in absolute value) for larger radii.}
  \label{fig:current_vs_f}
\end{figure}

The splitting effect can be  tuned by the value of either the amplitude of the periodic forcing or the static force.
Fig. \ref{fig:current_vs_f} represents the average value of the velocity {\it vs.} $f_0$ for the same geometry and $F_0 = 20$. One can see that by tuning $f_0$, one can control the separation of particles of different sizes. For instance, by selecting $f_{0}=-1.5$, small particles of radius $r=0.1b$ will move to the left with a velocity $-0.7$, whereas large particles of radius $r=0.9b$ will drift to the right with velocity $0.5$. More importantly, the velocity depends almost linearly on $f_0$, thus facilitating an efficient control of the separation effect. In addition,  by progressively changing $f_0$, one obtains a device that, with a fixed geometry, can be used to continuously separate particles of any size.

We have also analyzed the effect of the frequency of the periodic forcing on the particles current.
For small frequencies the results agree with those obtained in the
adiabatic limit. As the frequency increases the velocity becomes
progressively smaller. Eventually, at very high frequencies, the
change in the oscillating force is so fast that the particle cannot
follow it and as a result there is a vanishing effect of the oscillating force.

The fact that particles of different sizes travel in opposite directions leads to an efficient separation that can be improved further by increasing the number of periodic cavities. A way to quantify this is to calculate the probability that 
a large particle with an average positive velocity $v>0$ reaches the
``wrong'' collector, i.e. the left boundary of the device placed at
$x=nL$, where $n$ is a negative integer (see
Fig. \ref{fig:splitter}). This probability is
$P(x<nL,t)=0.5+ 0.5 \erf\left((nL-vt)/\sqrt{4D_{\textrm{eff}}t}\right)$
where we have assumed that the process can be described  as
a driven-diffusion process 
with an average velocity $v$ and an effective diffusion coefficient
$D_{\textrm{eff}}$. This probability peaks in
time at $t_{max}=-n/v$, and its maximum value is at
$P(x<nL,t_{max})=0.5+ 0.5 \erf\left(-\sqrt{v|n|/D_{\textrm{eff}}}\right)$.
We use this maximum probability as a measure of the purity of the sample collected at the end of the device.
For typical values of the velocity $v\sim0.2$ (cf. Fig.~\ref{fig:current_vs_f}) and $D_{eff}\lesssim1$,
one can achieve 99.9968\% of purity after only  $40$ periods.

Note, that the entropic splitting effect can be used for separation of
DNA fragments of different size using typical values as for DNA
electrophoresis. The DNA fragments are considered as randomly coiled
polymers with radius $R=d(M/M_0)^{1/2}$ where $M$ is the number of
base-pairs, $d=100nm$ is the Kuhn length, $M_0=300$ is the number of
base pairs per Kuhn length \cite{sorting}. Assuming Stokes' law and the
Sutherland-Einstein relation, (i) the friction coefficient for the dynamics of such a DNA fragment in water
is $\eta \approx 2\cdot 10^{-9}\,  \mathrm{kg/s} \,
\sqrt{ M/M_{0}} $ and (ii) the diffusion coefficient is
$D \approx 2\cdot 10^{-8}\, \mathrm{cm}^{2}\mathrm{/s} \,
\sqrt{ M_{0}/M }$.

For a device with $L= 3 \mu m$ and the same geometry and parameters as
in Fig.\ref{fig:channel}, typical values of the velocities are
 $L/\tau=D_0/L\sim 0.9 \mu m/s$, achievable by
electrophoresis with an electric field of $7\,\, V/cm$. For DNA chains of
$R\sim250 nm$ differing in radius by a $\sim25\%$, a purity of
separation of 99.997\%  results after $40$ periods.
These values compare favorably with commonly used techniques. Much
larger velocities and efficiencies are possible with smaller DNA
chains, or nanosized particles.

The effectiveness of the entropic splitting can  be tuned by
choosing  the geometry of the channel. In particular, the
entropic splitting effect  becomes more important upon increasing the
asymmetry of the walls, the slopes of the channel or by decreasing the
bottleneck width $b$. Altering the design of the channel geometry it is feasible to separate particles of very similar radii,
e.g. by choosing $b$ close to $r$. Moreover, yet another advantage of this set up is
that it works in a time-continuous mode.
%

In summary, we have presented a novel, purely entropic particle
splitting mechanism which is able to separate particles of different
sizes. The mechanism is based on the presence of an entropic
rectification of fluctuations caused by the  asymmetric
form of the channel. This rectification may overcome the effect of an applied force by reversing the motion of the particles. The mechanism is very efficient and can be controlled by  tuning the geometric parameters of the channel leading to both, different velocities and directions. This idea could be implemented in constrained structures with narrow channels and pores where entropic effects are important.

This work has been supported by the Icrea Academia Program, by the MICINN of the Spanish government through the I3 Program
and grant No. FIS2008-01299, by the Max Planck society, the Volkswagen
foundation project I/83902 and the German excellence cluster
''Nanosystems Initiative Munich'' (NIM).

\end{document}